\documentclass[letter, 12pt]{article}

\usepackage[utf8]{inputenc}

\usepackage{tabularx,booktabs,colortbl}
\usepackage{amsmath,array,graphicx}
\usepackage{tikz}
\usepackage{textcomp, gensymb}
\usepackage{float}
\usepackage{textgreek}
\usepackage{multicol}
\usepackage{titling}

\usepackage[super]{natbib}
\usepackage{amssymb}
\usepackage{makecell}
\usepackage{color,soul}
\usepackage[labelfont=bf]{caption}
\usepackage[margin=0.75in]{geometry}
\usepackage{subcaption}
\DeclareCaptionFormat{suboverlay}{\gdef\subcapoverlay{(\thesubfigure) #3\par}}
\DeclareCaptionStyle{suboverlay}{format=suboverlay}

\newcommand{\subcaptionOverlay}[1]{%
  \subcaption{}%
  \begin{tikzpicture}
    \node [inner sep=0,anchor=north west]at (-3ex,3ex) (image) {#1};
    \draw node [black] {\subcapoverlay};
  \end{tikzpicture}%
}
\renewcommand\hl[1]{#1} 

\title{Atomistic mechanisms underlying the maximum in diffusivity in doped Li$_7$La$_3$Zr$_2$O$_{12}$}

\author{Juan C. Verduzco, Ernesto E. Marinero, and Alejandro Strachan*
\\
*Corresponding author: \\
Alejandro Strachan (strachan@purdue.edu)\\
Phone: +1 765 496 3551 \\
Fax: +1 765 494 1204 
}

\date{School of Materials Engineering and Birck Nanotechnology Center \\
Purdue University, West Lafayette, Indiana, 47907 USA \\
}

\begin{document}

\maketitle

\begin{abstract}

Doped lithium lanthanum zirconium oxide (LLZO) is a promising class of solid electrolytes for lithium\hl{-}ion batteries due to their good electrochemical stability and compatibility with Li metal anodes. Ionic diffusivity in these ceramics is known to occur via correlated, vacancy mediated, jumps of Li+ between alternating tetrahedral and octahedral sites. Aliovalent doping at the Zr-site increases the concentration of vacancies in the Li+ sublattice and cation diffusivity\hl{,} but such an increase is universally followed by a decrease for Li+ concentration lower than 6.3 - 6.5 Li molar content. Molecular dynamics simulations based on density functional theory show that the maximum in diffusivity originates from competing effects between the increased vacancy concentration and the increasing occupancy of the low-energy tetrahedral sites by Li+\hl{, which} increases the overall activation energy associated with diffusion. For the relatively high temperatures of our simulations, Li+ concentration plays a dominant role in transport as compared to dopant chemistry.
\end{abstract}

\section{Introduction}

Lithium batteries with high energy density, improved safety, and faster charging rates are instrumental \hl{in meeting} the energy demands for electric vehicles \cite{duan2020building}, advanced wearable and portable devices \cite{fan2020flexible}, and renewable energy storage systems \cite{diouf2015potential}. Conventional liquid electrolyte based batteries cannot satisfy these requirements\hl{,} as they are incompatible with high\hl{-}energy anode materials and fall short in key safety features \cite{fan2018recent}. Replacing organic liquid electrolytes with solid electrolytes in these systems is expected to improve mechanical properties, minimize safety risks due to improved thermal and electrochemical stability, and enable the pairing with high\hl{-}energy\hl{-}density anodes to improve battery capacity. However, practical applications for solid-state batteries are limited, as liquid chemistries \hl{generally} surprass solid electrolytes in important metrics such as interfacial resistance and ionic conductivity \cite{fan2018recent, li2021advance}.

Lithium lanthanum zirconium oxide (Li$_7$La$_3$Zr$_2$O$_{12}$, LLZO) ceramic garnet has \hl{emerged as} a prominent and promising candidate for all-solid-state battery applications \cite{murugan2007fast}. Aliovalent dopants provide an important avenue for the optimiz\hl{ing} its transport properties \cite{samson2019bird}. In LLZO, lithium ions occupy alternating tetrahedral (24d) and a distorted octahedral (96h) sites \hl{forming} 3D channels for ionic diffusion. As is often the case in solid\hl{-}state diffusion, transport in the Li+ sublattice is mediated  by vacancies \cite{han2012experimental}.  LLZO exhibits room temperature ionic conductivity of \hl{approximately} $ 1 \times 10^{-4}$ S/cm \cite{murugan2007fast}, while doped LLZO has been reported to \hl{reach roughly} $~1.8 \times 10^{-3}$ S/cm for a dual\hl{-}substituted Ga/Sc ceramic \cite{buannic2017dual}.

Substitutional dopants have multiple effects on LLZO\hl{,} and several mechanisms have been proposed to explain the role of doping on its ionic conductivity. Doping affects the relative stability between two polymorphs and stabilization of the cubic polymorph, \hl{which has} ionic conductivity two orders of magnitude higher than the tetragonal phase, is critical \mbox{\cite{awaka2009synthesis, thompson2014tetragonal}}. However, doping has additional effects and can significantly \hl{influence} transport within the cubic phase. Aliovalent dopants alter the occupancy of the Li-sublattice since charge compensation mainly occurs through the introduction of Li vacancies \cite{squires2019native}. The increase in vacancies accelerates diffusion but reduces the number of carriers in the sublattice. Dopants also alter the local chemistry experienced by the mobile Li+ ions during diffusion. Changes in the 3D channels of Li-sublattice can introduce bottlenecks that alter Li migration pathways \cite{zhang2018regulation}. Experimental results across different dopants show that while the detailed chemistry affects transport, a general trend \hl{emerges} of increasing conductivity with decreasing Li+ content for weak doping\hl{,} followed by a maximum between 6.3 – 6.5 Li molar content\hl{,} and a subsequent reduction \cite{thompson2014tetragonal, zeier2014structural, schwanz2020bi, verduzco2021active}.  

To elucidate the mechanisms responsible for the experimental observations, several computational studies \hl{have utilized} electronic structure calculations \hl{to }investigate transport in tetragonal and cubic LLZO \cite{bernstein2012origin, meier2014solid}\hl{as well as} the energetics of doping and its effect on transport \cite{jalem2013concerted, miara2013effect, jalem2015insights}. Dopants can occupy various sites in the LLZO structure. Miara et al. calculated defect energies to map the site and oxidation state preference for a \hl{broad} range of elements, \hl{demonstrating the} potential for many stable doped LLZO structures \cite{miara2015first}. 

Molecular dynamics (MD) simulations based on density functional theory (DFT) have \hl{demonstrated} that correlated motions of Li ions are a key mechanism behind the superionic conductivity of cubic LLZO ceramics \cite{jalem2013concerted}. These simulations \hl{have revealed} that transport within the Li+ sublattice is not dominated by independent hops between alternating sites\hl{,} but by multiple ions experiencing collective transitions (correlated  in time and space)\cite{chen2017data}. In contrast, Meier et al. \cite{meier2014solid} suggest mechanisms characterized by asynchronous jumps in cubic LLZO\hl{, while} Chen et al. \cite{chen2017data} found that diffusion in cubic LLZO is hindered by back-and-forth jumps, originat\hl{ing} from the low concentration of vacancies in LLZO. Monte Carlo simulations on the garnet lattice provided insight and established general trends on the effect of Li+ concentration, the relative energy between sites, and short range interactions on transport \cite{morgan2017lattice}, but do not capture \hl{specific} chemical effects or long range electrostatic interactions important in the cases of interest here. Thus, important questions regarding the underlying transport mechanisms in LLZO and \hl{crucially}, the observed maximum in ionic diffusitity vs. Li+ concentration observed across dopants\hl{,} remains unexplained.

In this paper, we analyze how changes in Li+ concentration\hl{, resulting} from aliovalent substitution and by artificially removing Li atoms in the simulation\hl{,} affect the population of octahedral (O) sites and tetrahedral (T) sites, associated residence times, and correlations between the various individual Li+ jumps. We find that the maximum diffusivity with decreasing Li+ originates from the competition between increasing vacancies that accelerates transport and the increasing occupancy of low-energy T sites that results in an increase in the overall activation energy for diffusion. \hl{Furthermore,} our simulations, at temperatures above 1200K, show correlated hops within timescales of a picosecond. \hl{T}hese correlations decrease with decreasing temperature\hl{,} and reveal how doping influences correlated jumps.

\section{Computational Details}

MD simulations were \hl{executed} using Kohn-Sham DFT as implemented in the Vienna Ab-initio Simulation Package (VASP) \cite{kresseEfficiencyAbinitioTotal1996, kresseEfficientIterativeSchemes1996}. The chosen kinetic energy cutoff is a compromise between accuracy and computational expedience, given the large number of MD simulations required to explore chemistry and compositional space. The simulations reported in this paper required over 1.7 million core hours of computing. \hl{To verify} the accuracy of the results, we \hl{conducted} two additional simulations using a kinetic energy cutoff of 500 eV and \hl{can} confirm \hl{both} qualitative and quantitative \hl{agreement with} the results with the smaller basis set. The exchange and correlation functional were described within the generalized gradient approximation using PBEsol \cite{perdew2008restoring}, \hl{which} improves the Perdew-Burke-Ernzernhof (PBE) \cite{perdewGeneralizedGradientApproximation1996} functional for solids. We used projector augmented wave (PAW) pseudopotentials to describe core electrons \cite{blochlProjectorAugmentedwaveMethod1994}. No spin polarization was considered.

The crystal structure for cubic LLZO with formula $Li_{56}La_{24}Zr_{16}O_{96}$ \cite{chen2015origin} was obtained from the \hl{I}norganic \hl{C}rystal \hl{S}tructure \hl{D}atabase (ICSD) \cite{bergerhoff1987crystallographic}. To generate Bi-doped and Ta-doped structures, we stochastically replaced the desired number of Zr atoms with Bi/Ta. Compositions analyzed were created with substitutions in the formula $Li_{56-x}La_{24}Zr_{16-x}O_{96}$ for x = (4, 6, 8) to mirror compositions $Li_{7-x}La_{3}Zr_{2-x}M_{x}O_{12}$. Substitution of the Zr 16a site by Bi and Ta has been reported in prior experiments \cite{wagner2016synthesis, schwanz2020bi, logeat2012order}. Due to the supervalent substitution, \hl{one} lithium atom is removed per Bi or Ta, to account for the change in oxidation state \hl{from} Zr (4+) \hl{to} Tantalum/Bismuth (5+). We studied compositions between Li molar contents, $x_{Li}$, 0 and 1 $Li_{7-x}La_{3}Zr_{2-x}M_{x}O_{12}$ for x = (0, 0.5, 0.75, 1), resulting in a range of overall occupancy of the Li+ sublattice from 48/72 to 56/72. To separate the role of chemistry and the change in Li lattice occupancy on transport, we performed simulations on structures where Li+ cations \hl{were} removed without the substitutional doping. 
\hl{Since} all these relatively small systems can be influenced by the different atomistic configurations resulting from aliovalent substitutions\hl{, each} structure was independently generated with stochastic replacements, thus exploring distinct arrangements of the local environments. Lattice parameters were taken from room temperature experimental measurements for pure LLZO, Bi-doped LLZO \cite{wagner2016synthesis}\hl{,} and Ta-doped LLZO \cite{wang2012high} as shown in Table \ref{latticetable}.

\begin{table}[ht]
\caption{Experimental lattice parameters for doped LLZO structures.}
\label{latticetable}
\vskip 0.15 in
\begin{center}
\begin{small}
\begin{sc}
\begin{tabular}{lccccr}
\toprule
Structure	&	\makecell{Lattice \\ Parameter}	&	\makecell{Number of \\ Atoms}	& \makecell{Number of \\ Lithium atoms}	\\
\midrule
LLZO	        &	12.994 Å	&	192	 &	56	\\
LLZO (Bi = 0.5)	&	13.019 Å	&	188	 &	52	\\
LLZO (Bi = 0.75)&	13.031 Å	&	186  &	50 \\
LLZO (Bi = 1)	&	13.045 Å	&	184  &	48 \\
LLZO (Ta = 0.5)	&	12.919 Å	&	188	 &	52 \\
LLZO (Ta = 0.75)&	12.904 Å	&	186	 &	50 \\
LLZO (Ta = 1)	&	12.890 Å	&	184	 &	40	\\
LLZO (LiV = 0.5)	&	12.994 Å	&	188	 &	52	\\
LLZO (LiV = 0.75)	&	12.994 Å	&	186	 &	50	\\
LLZO (LiV = 1)	&	12.994 Å	&	184	 &	48	\\
LLZO (LiV = 1.25)	&	12.994 Å	&	182	 &	46	\\
\bottomrule
\end{tabular}
\end{sc}
\end{small}
\end{center}
\vskip -0.1in
\end{table}

As mentioned \hl{previously}, the Li sublattice consists of two types of sites\hl{:} 24d T sites and 96h O sites. The latter comprise\hl{s} closely located pairs\hl{,} and previous studies have shown that two lithium atoms cannot simultaneously occupy the\hl{se} split 96h sites \cite{miara2013effect}. This effectively leaves us with only 48 possible O spaces for \hl{ion diffusion}. Given that tetrahedral sites are known to be energetically favorable in similar structures \cite{wang2015design,he2017origin}, we initially distributed the Li atoms to occupy all the tetrahedral sites\hl{, with} the remaining atoms randomly placed on the octahedral split sites 96h\hl{, ensuring} that no consecutive split sites were simultaneously occupied. During the MD equilibration of the structures, atoms migrated from this initial configuration to preferentially occupy the O sites\hl{,} driven \hl{by} the minimization of electrostatic repulsion between Li+ ions\hl{,}in good agreement with prior work.

We performed isochoric and isothermal MD simulations at T = 1273 K and T = 1773K with a 1 fs timestep. These temperatures were selected to capture diffusive behavior at timescales achievable with DFT-based MD. All molecular dynamics simulations ran \hl{for} between 40 and 60 ps, long enough to achieve the diffusive regime and extract a diffusion constant\hl{,} as demonstrated by the MSD plots in Fig. S1. A Nos\'{e}-Hoover thermostat was used to sample the canonical ensemble with a mass corresponding to a period of 40fs. We note that due to the random nature of the doping, the results can be affected by the specific configuration created. Each structure was \hl{independently and stochastically generated, allowing us to assess} the effect\hl{s} of atomic structure and other simulation details on transport and other properties by exploring trends across chemistries and compositions. As with all physics-based modeling, our simulations are not without approximations. We did not include relativistic effects, which could play a role in our systems (especially those with Bi), for consistency across the various chemistries and to \hl{keep computational resources manageable}. In addition, we artificially maintained the neutrality of the system with artificial vacancies (i.e. we removed Li atoms not Li+). The motivation behind this choice is the challenge associated with the presence of several localized charges in cells with periodic boundary conditions. These approximations are justified a posteriori by our finding of similar overall trends in terms of diffusivity and site occupancy between Ta and Bi doping and the case with artificial vacancies. Future work to explore the effect of these assumptions would be of interest.

\section{Results \& Discussion}
\subsection{Diffusivity vs. Li+ concentration}

\begin{figure*}[ht]
\centering
\begin{subfigure}{.49\textwidth}
\centering
\subcaptionOverlay{\includegraphics[width=\textwidth]{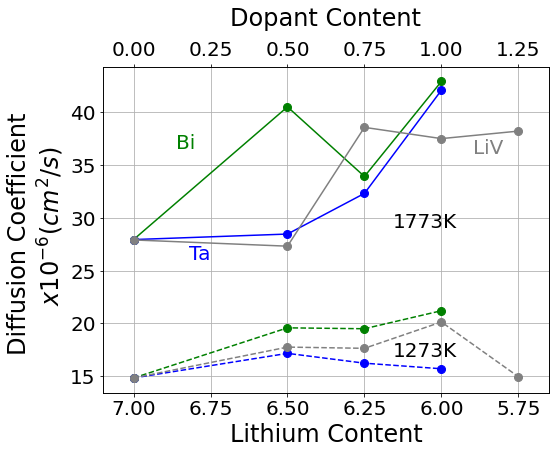}}
\end{subfigure}
\begin{subfigure}{.50\textwidth}
\centering
\subcaptionOverlay{\includegraphics[width=\textwidth]{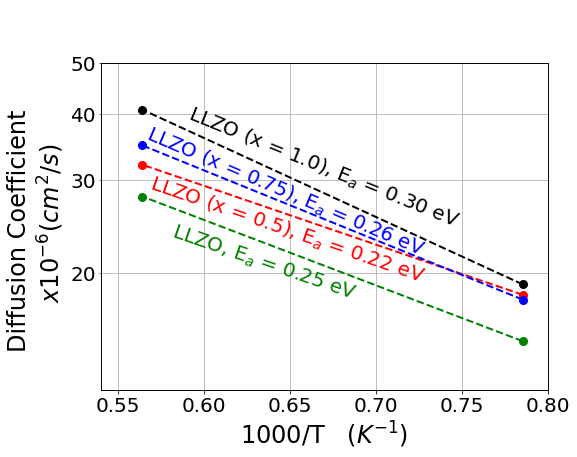}}
\end{subfigure}
\caption{(a) Diffusion coefficients versus Lithium content. Colors represent different dopants, and LiV represents a structure created with artificial vacancies.(b) Diffusion coefficients versus inverse temperature. Colors indicate different dopant contents. 
}
\label{msdcoeffarrh}
\end{figure*}

To compute the diffusion constant of Li+, we calculated the mean squared displacement ($<R^2>$) of Li+ as a function of time (see SI, \hl{Fig.} \ref{rawMSD}) and fitted the linear section (after a diffusive regime is achieved) to the expression $<R^2>=3Dt$. \hl{Fig.} \ref{msdcoeffarrh}(a) shows the diffusion coefficient obtained from our MD simulations as a function of lithium content (dopant quantity on \hl{the} top axis) for both temperatures and the two dopants studied\hl{,} as well as the models with artificially modified Li contents. \hl{Interestingly, our results suggest} that the occupancy of the Li sublattice \hl{has a more significant influence on} Li+ transport \hl{than} the specific chemistry of the dopant.

At T = 1773 K, we find a \hl{monotonic} increase \hl{in} diffusivity with decreasing Li+ content for all materials studied. Quite interestingly, at T = 1273 K\hl{,} we observe an initial increase in conductivity with doping, a maximum at an intermediate Li+ concentration of approximately 6.5, followed by either a reduction or plateau. This is consistent with a maximum in ionic conductivity for a Li+ molar content of approximately 6.5 observed experimentally at lower temperatures \cite{thompson2014tetragonal, zeier2014structural, verduzco2021active}. We attribute deviations from these general trends to the small sample sizes and short simulation time\hl{s} intrinsic to the DFT-MD simulations. Accurate interatomic potentials for these systems, including dopants, would enable \hl{simulations of} larger systems and longer timescales\hl{,} which would reduce the noise in the diffusivity results. \hl{Fig.} \ref{msdcoeffarrh}(b) shows an Arrhenius plot of the resulting diffusion coefficients and estimates of the effective activation energies. These activation energies are rough estimates and \hl{are} computed to confirm independent observations regarding the mechanisms behind the maximum in conductivity. The remainder of the paper analyzes the mechanisms behind the trends observed in \hl{Fig.} \ref{msdcoeffarrh}. 

\begin{figure*}[ht]
    \centering
    \includegraphics[width=0.5\textwidth]{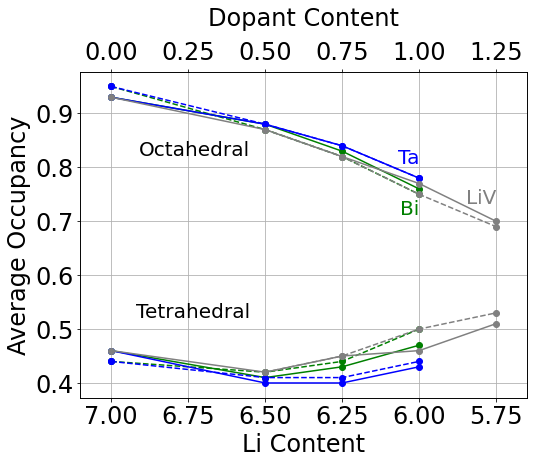}
    \caption{Site-wise occupancy of different doped-LLZO structures from MD simulations. Colors represent different dopants, and LiV represents a structure created with artificial vacancies. Dashed lines represent T=1273 K and solid lines T=1773 K. The two groups of lines describe the two independent sites in the Li sublattice.}
    \label{occupancies}
\end{figure*}

The Li+ sublattice consists of 24 tetrahedral sites and 48 split-octahedral sites, and transport \hl{occurs} via alternating jumps between tetrahedral and octahedral sites \cite{jalem2013concerted, adams2012ion}. To understand the trends in diffusivity discussed above\hl{,} we performed a detailed analysis of the MD trajectories\hl{,} tracking the time evolution of individual Li+ ions. This \hl{was} done by mapping each ion \hl{o}nto a sublattice site using a distance-based criterion and recording each jump. \hl{Fig.} \ref{occupancies} shows the average occupancy of the O and T sites, $n_O$ and $n_T$, as a function of Li+ content. Confidence intervals (95\%) for these values are estimated at 0.003 for tetrahedral occupancy and 0.001 for octahedral occupancy. As noted in prior work \cite{meier2014solid, he2017origin}, in LLZO the Li+ ions preferentially occupy the more populous, but higher energy, O sites to minimize electrostatic repulsion. A reduction of the Li+ concentration (either by doping or artificially) from the LLZO value results in an initial reduction in both $n_O$ and $n_T$. However, for Li+ molar contents below 6.5\hl{,} the population of the low-energy T sites actually increases at \hl{the} expense of O sites \cite{adams2012ion}. These observations apply regardless of temperature and dopant chemistry. As discussed in Section 2, we performed two additional simulations with a more complete basis set. They correspond to stoichiometric LLZO and LLZO (LiV = 1.25). After 25 ps of DFT\hl{-}MD, the values for site occupancy are within 5\% of those reported in Fig. \mbox{\ref{occupancies}} and the residence times are within 15\%. The increase in $n_T$ with decreasing Li+ content might appear counterintuitive at first sight\hl{,} but it is the result of the reduced importance of electrostatic repulsion (which drives the ions to the O sites in pure LLZO)\hl{, with} ions preferring the low-energy T sites. We note that Monte Carlo simulations \cite{morgan2017lattice} including short\hl{-}range interactions do not capture the minimum in occupancy of the T sites in \hl{the} range of interest for doped LLZO observed here.

Having established where Li+ ions are located as a function of Li+ molar content, we \hl{now} focus on the jumps between neighboring sites. Since the network topology forces ions to move by alternating \hl{between} O and T sites, the number of T$\rightarrow$O and O$\rightarrow$T jumps over time tend to the same value and are nearly identical in our simulations (see SI, \hl{Fig.} \ref{transitiontable}). Thus, the average residence times, $\tau_O$ and $\tau_T$, of the Li+ atoms in O and T sites respectively, are related to the occupancies according to:

\begin{equation}
\frac{24*n_T}{\tau_T} = \frac{48*n_O}{\tau_O}
\label{eq1}
\end{equation}

where the coefficients 24 and 48 represent the number of T and O sites in the unit cell\hl{,} respectively. We note that transitions between octahedral sites (96h) that bypass the tetrahedral site have been reported for garnets with lower Li contents \mbox{\cite{xu2012mechanisms}}. However, such processes \hl{were not observed in our study, which} is consistent with previous reports \mbox{\cite{han2012experimental, jalem2013concerted, wang2014local}}.

\begin{figure*}[ht]
\centering
\begin{subfigure}{.49\textwidth}
  \centering
\subcaptionOverlay{\includegraphics[width=\textwidth]{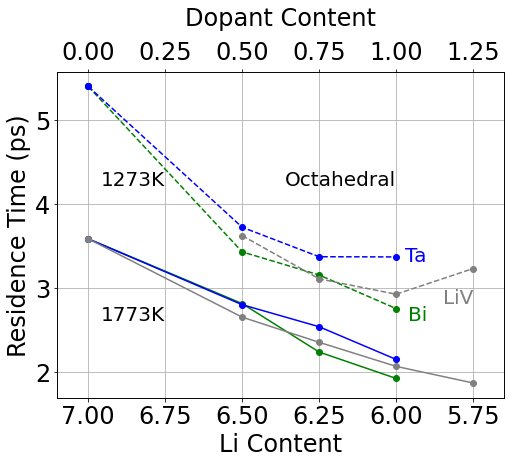}}
\end{subfigure}%
\begin{subfigure}{.51\textwidth}
  \centering
\subcaptionOverlay{\includegraphics[width=\textwidth]{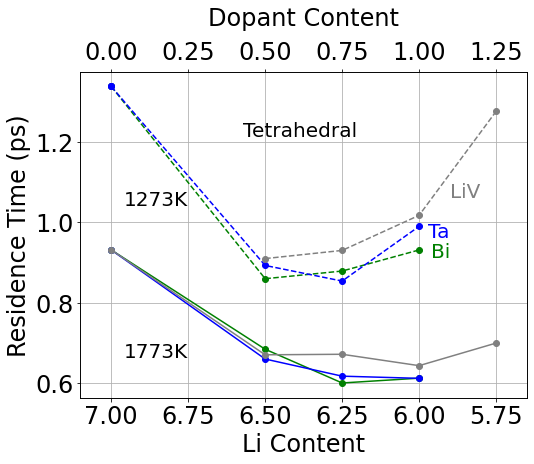}}
\end{subfigure}
\caption{Average residence time of different doped-LLZO structures from MD simulations against the stoichiometric Li molar content, (a) for O sites and (b) for T sites.}
\label{residtime}
\end{figure*}

The computed average residence times in O and T sites are shown as a function of Li content in \hl{Fig.} \ref{residtime}. As expected from Eq. \ref{eq1}, the residence times in O sites are longer than those for T sites. Jumping statistics reveal a Poisson-like diffusion behavior for these systems, in agreement with previous reports for undoped cubic LLZO garnets \mbox{\cite{chen2017data}}. Representative residence time distributions and confidence intervals (95\%) calculated from our analysis are included as \hl{Fig.} \mbox{\ref{jumpstats_figs3}} and Table \mbox{\ref{rtime_statisticstable}}. At T = 1773 K\hl{,} the residence time in O sites decreases monotonically with decreasing Li+ concentration, while the residence time in T sites follows a similar trend for low doping but levels off for Li+ molar content below 6.5. This plateau is due to the increase in occupancy of T sites ($n_T$) as shown in \hl{Fig.} \ref{occupancies}. Faster diffusion and shorter residence times are expected since reducing the Li+ concentration increases the number of vacant sites and the number of available jumps at any given time. At T = 1273 K, we observe similar trends for Li+ concentrations near the LLZO values\hl{. However}, remarkably, as $x_{Li}$ decreases below approximately 6.5\hl{,} the residence time for the O sites plateaus and that for T sites increases quite sharply. This is associated with the maximum in diffusion coefficient and its subsequent decline with doping for Li+ molar contents below 6.5. 

The relative values of $\tau_O$ and $\tau_T$ are a direct consequence of the increasing occupation of T sites and Eq. \ref{eq1}\hl{,} but the trend in absolute residence times observed at 1273 K, which correlates with the observed maximum in diffusivity, is more subtle. The initial decrease in residence times is due to the increase in the density of vacancies. The increase in residence time and decrease in diffusion coefficient at T = 1273K for $x_{Li}$ below \hl{approximately} 6.5 can be understood by recalling that the T sites are lower in energy than the O sites for Li+. Thus, the topologically-required increase in the fraction of T$\rightarrow$O jumps with increasing population of (low energy) T sites results in a higher overall activation energy for diffusion. Our MD results confirm this picture\hl{. An} effective activation energy can be estimated from the Arrhenius plots in \hl{Fig.} \ref{msdcoeffarrh}(b) that indicate a lower activation energy for $Li_{7-x}La_{3}Zr_{2-x}M_{x}O_{12}$ with x = 0.5. 

\subsection{Correlated Events and Li+ transport}

\begin{figure*}[ht]
\centering
\begin{subfigure}{.505\textwidth}
  \centering
\subcaptionOverlay{\includegraphics[width=\textwidth]{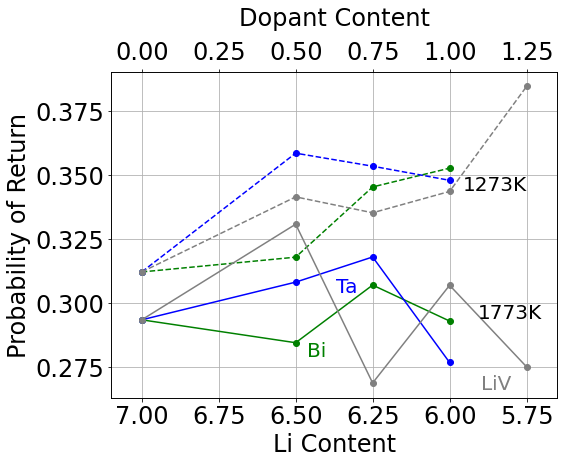}}
\end{subfigure}%
\begin{subfigure}{.495\textwidth}
  \centering
\subcaptionOverlay{\includegraphics[width=\textwidth]{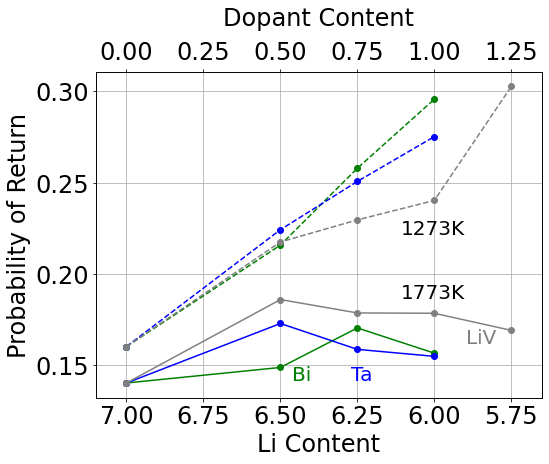}}
\end{subfigure}
\caption{Return probabilities for doped-LLZO structures from MD simulations against the stoichiometric Li molar content, (a) for T$\rightarrow$O$\rightarrow$T jumps, and (b) O$\rightarrow$T$\rightarrow$O jumps. Colors represent different dopants, and LiV represents a structure created with artificial vacancies. Dashed lines represent T=1273 K and solid lines T=1773 K.}
\label{returnprobs}
\end{figure*}

The residence times discussed above capture the overall trends of diffusivity vs. composition but do not account for the correlation between neighboring events\hl{,} which has been shown to play an important role in ionic transport in LLZO \cite{meier2014solid, chen2017data}. This section quantifies the percentage of return jumps, when two subsequent jumps leave an ion in its original site, and time correlations between neighboring events. Taking advantage of the clearly defined 3D channels of the LLZO structure, and the alternating pattern of T and O sites, we computed the fraction of ions jumping back to their original site after a transition. An ideal random walk in the Li+ sublattice would have a return probability of 50\% for T$\rightarrow$O$\rightarrow$T (since each O site is connected to two T sites) and 25\% for O$\rightarrow$T$\rightarrow$O transitions. \hl{Fig.} \ref{returnprobs} shows return probabilities as a function of Li+ concentration for the various cases studied. To quantify the uncertainty of these probabilities, an ensemble approach subdividing the simulation shows an average deviation from the mean of $\pm$ 0.025. As with other transport characteristics, chemistry plays a minor role as compared with Li+ concentration for the conditions studied. We find that the return probability following a T$\rightarrow$O jump ranges from 0.25 to 0.35, significantly below the random value, indicating a higher than random chance of moving forward. The simulations reveal an overall trend towards the random value with decreasing Li+ content at T = 1273 K and little change at T = 1773 K. In the case of O$\rightarrow$T transitions, we also observe an increase in return fractions with doping at the lower temperature and flat value at 1773 K\hl{,} but the numbers are closer to the random case (actually\hl{,} our results show values over 0.25 for the low temperature and smallest Li+ concentrations).

The return probabilities are consistent with prior results indicating concerted migration \cite{jalem2013concerted, jalem2015insights, chen2017data, he2017origin} where several atoms move nearly simultaneously. As will be shown below, vacating a T site often results in nearly instantaneous occupation by a neighboring Li+, preventing a return. The observation of return probabilities closer to the random value for O$\rightarrow$T can be attributed to the lower average occupancy of the T site\hl{,} which reduces the probability of a neighboring Li+ occupying the vacated O site. Our results differ from those in Ref. \cite{chen2017data}\hl{,} which reported significant return jumps attributed to low vacancy concentrations. The origin of this difference is unclear at this point\hl{;} it could be due the \hl{fact} that our work uses DFT simulations and Chen et al. \hl{use} a force field or differences in the analysis. We note that less-than-random return probabilities are consistent with the widely accepted correlated hops involving multiple ions.

\begin{figure*}[ht]
    \centering
    \includegraphics[width=\textwidth]{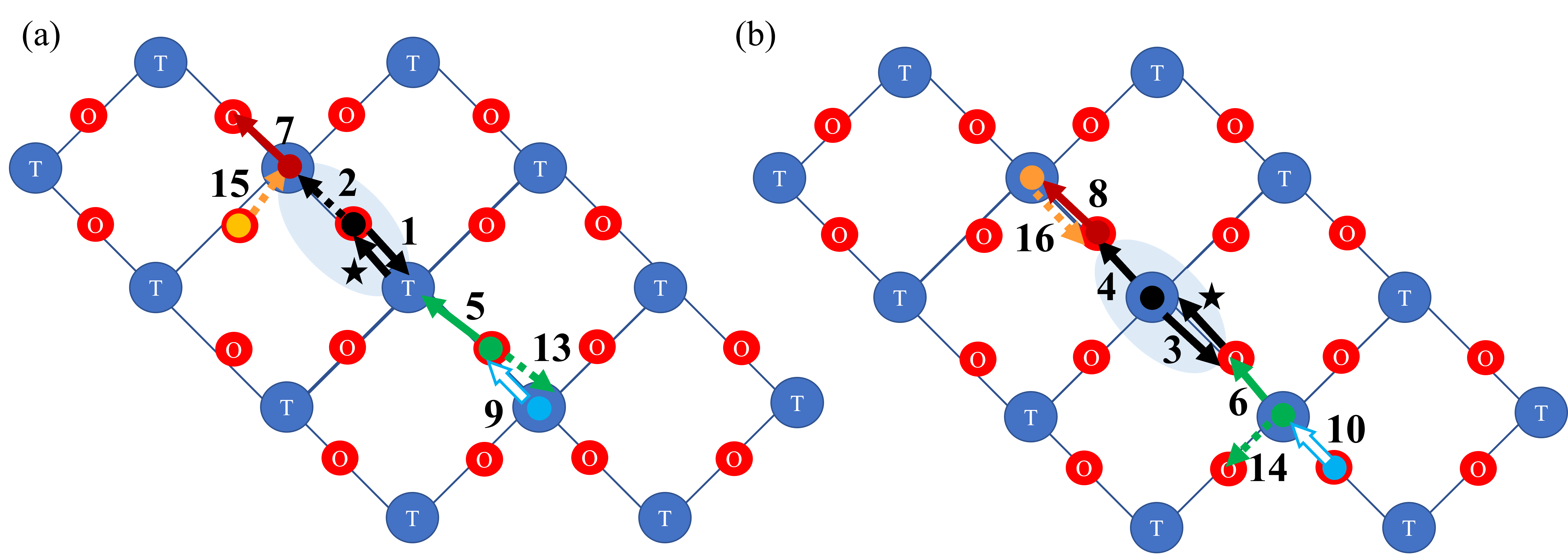}
    \caption{Schematic representation of possible secondary events to study correlated transitions. Initial jumps are marked with a ($\bigstar$). Colors represent different Li atoms occupying different sites. Arrows represent the rate of secondary transitions, with dashed lines being anticorrelated events.  Arrows representing multiplicity of events were omitted for clarity. (a) shows \hl{Cases} with an initial event as T$\rightarrow$O and (b) O$\rightarrow$T }
    \label{correlatedschematic}
\end{figure*}

To gain \hl{additional} insight into correlated jumps, we characterized time correlations between jumps in neighboring sites. We computed the distribution of times between several pair of events, labeled from 1 to 16, see Fig. \ref{correlatedschematic}. The first event\hl{,} which can be a T$\rightarrow$O transition (\hl{Fig.} \ref{correlatedschematic}(a) or a O$\rightarrow$T (\hl{Fig.} \ref{correlatedschematic}(b)) \hl{transition is} shown as \hl{a} black arrow and marked with a star. The second event could involve the same Li+ (black arrows) or a different ion \hl{(color arrow). The} number next to the arrow labels the event (see also Table \ref{pairwisetable} in the SI). We define time correlation functions between two events, $E_1$ at time $t_1$ and $E_2$ at time $t_2$ as:
\begin{equation}
C_{E_1E_2}(t) = \frac{n_{E_1E_2}(t,\Delta)}{N_{E_1} \cdot n_{E_2|E_1}(\Delta) } 
\end{equation}
where the numerator represents the number of $E_2$ events following a $E_1$ event after a time between $t$ and $t+\Delta$ normalized by $\Delta$. The denominator is a normalization factor obtained as the product of number of $E_1$ times the average rate of event $E_2$ given the configuration resulting from $E_1$.

\begin{figure}[H]
    \centering
    \includegraphics[width=\textwidth]{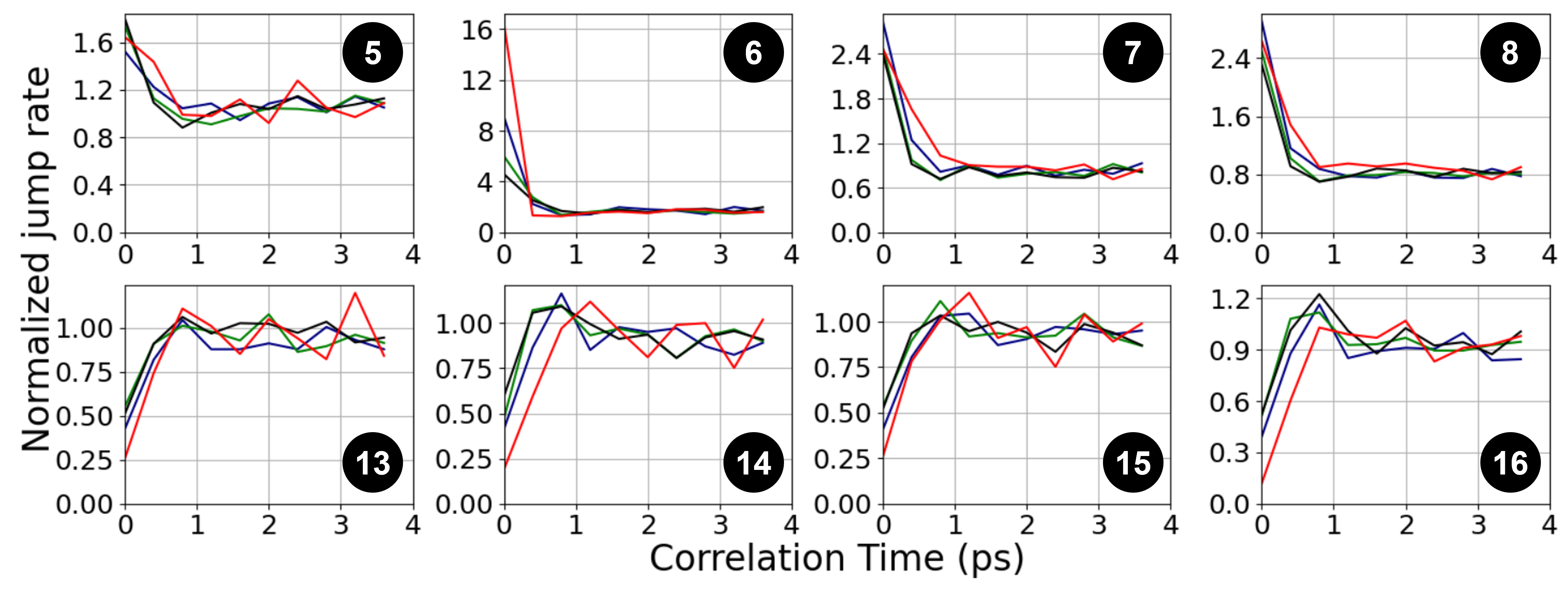}
    \caption{Normalized jumping rate against correlation time for selected \hl{Cases} for the family of Bi-doped LLZO structures at 1773K. Colors represent different dopant molar contents. [Red = 0, Blue = 0.5, Green = 0.75, Black = 1]. Table \ref{pairwisetable} describes of the pair correlation cases.}
    \label{norm_corr_plots}
\end{figure}

\begin{minipage}{\textwidth}
\begin{table}[H]
\caption{Summary of pair-wise correlated mechanisms studied. T1 and T2 represent the first and second tetrahedral sites involved. O1 and O2 represent the first and second octahedral sites involved.}
\label{pairwisetable}
\vskip 0.15 in
\begin{center}
\begin{small}
\begin{sc}
\begin{tabular}{lcccccr}
\toprule
Case	&	Same Li	&	\makecell{First \\ Event}	&	\makecell{Second \\ Event}	&	Comment	&	Multiplicity	\\
\midrule
1	&	Yes	&	T1$\rightarrow$O1 	&	O1$\rightarrow$T1	&		&		\\
2	&	Yes	&	T1$\rightarrow$O1 	&	O1$\rightarrow$T2	&		&		\\
3	&	Yes	&	O1$\rightarrow$T1 	&	T1$\rightarrow$O1	&		&		\\
4	&	Yes	&	O1$\rightarrow$T1 	&	T1$\rightarrow$O2	&		&		\\
5	&	No	&	T1$\rightarrow$O1 	&	O2$\rightarrow$T1	&		&	3	\\
6	&	No	&	O1$\rightarrow$T1 	&	T2$\rightarrow$O1	&		&	1	\\
7	&	No	&	T1$\rightarrow$O1	&	T2$\rightarrow$O2 	&	O1 + T2 (1st NN)\footnote{1st NN indicates first nearest neighbors in the sublattice.}	&	3	\\
8	&	No	&	O1$\rightarrow$T1 	&	O2$\rightarrow$T2 	&	T1 + O2 (1st NN)	&	3	\\
9	&	No	&	T1$\rightarrow$O1	&	T2$\rightarrow$O2 	&	T1 + O2 (1st NN)	&	3	\\
10	&	No	&	O1$\rightarrow$T1 	&	O2$\rightarrow$T2 	&	O1 + T2 (1st NN)	&	3	\\
13	&	No	&	T1$\rightarrow$O1	&	O2$\rightarrow$T2 	&	T1 + O2 (1st NN) 	&	3	\\
14	&	No	&	O1$\rightarrow$T1 	&	T2$\rightarrow$O2 	&	O1 + T2 (1st NN) 	&	3	\\
15	&	No	&	T1$\rightarrow$O1	&	O2$\rightarrow$T2 	&	O1 + T2 (1st NN)	&	3	\\
16	&	No	&	O1$\rightarrow$T1 	&	T2$\rightarrow$O2 	&	T1 + O2 (1st NN) 	&	3	\\

\bottomrule
\end{tabular}
\end{sc}
\end{small}
\end{center}
\vskip -0.1in
\end{table}
\end{minipage}
\\
\\
\hl{Fig.} \ref{norm_corr_plots} shows the time correlations for key pairs of jumps for Bi-doped LLZO. Plots for the lower temperature time correlations are included as Fig. \ref{supp:lowtemp_norm_corr_plots}. Case numbers follow those in our analysis scripts shared on Github \cite{strachangroup-dopedllzo}. Cases 11 and 12 are not shown because they involve 3 Li atoms. The event pairs shown in the top row exhibit short-time correlations, i.e. $E_2$ is more likely to occur soon after $E_1$ while in the cases shown in the second row event $E_1$ inhibits $E_2$. In all cases, time correlations are negligible beyond 1 ps, with this correlation time being consistent with results in Ref. \cite{chen2017data}. As expected, in some cases, \hl{reducing} Li+ concentration reduces correlations. The strongest short-range correlation observed is case 6, where a vacated O site is filled from a neighboring T site. Interestingly, the related process associated with a vacated T site shows a much weaker time correlation. We attribute this to the electrostatic repulsion that tends to maximize the occupancy of O sites. Not surprisingly, the short-time correlation in case 6 weakens drastically with the reduction in Li+ content. \hl{Cases} 13 and 14, which track the opposite \hl{transitions} of \hl{Cases} 5 and 6, where the second atom does not follow the initial jump, are anticorrelated at short times. \hl{Cases} 7 and 8 describe time correlated motions where a site is vacated followed by a first-neighboring site being occupied (due to the reduction in \hl{electrostatic} repulsion) and \hl{C}ases 15 and 16 show, conversely, that the opposite motion is hindered. 

\section*{Conclusions}

In summary, \hl{DFT-MD simulations} reveal the mechanisms behind the role of aliovalent doping on ion transport in LLZO. Consistent with experiments, the simulations at T=1273 K on $Li_{7-x}La_{3}Zr_{2-x}M_{x}O_{12}$ with M=Ta/Bi show an increase in Li diffusion \hl{coefficient} with doping\hl{,} up to x$\sim$0.5 (Li+ concentration of \hl{approximately} 6.5)\hl{,} followed by a decrease. The mechanisms responsible for the maximum in diffusivity are the main focus of this paper. Analysis of the MD simulations reveals an initial decrease in the occupancy of both octahedral and tetrahedral sites with doping and the consequent reduction in Li+ concentration. Surprisingly, we find a change in this trend\hl{, with} an increase in the occupancy of the low-energy T sites with decreasing Li+ concentration. This is due to a reduction in the electrostatic repulsion that forces the Li+ to the high-energy O sites in pure LLZO. The maximum in diffusivity originates from the interplay between the increase in vacancy concentration with doping that speeds up transport with the higher occupation of the low-energy T sites that increases the overall activation energy for diffusion. A detailed analysis of the correlations between different jumps shows the importance of correlated motions and that correlations are seen not just between neighboring sites but also between second nearest neighbors. Our results indicate that increased Li+ conductivity would be achieved by reducing the energy difference between the octahedral and tetrahedral sites. 

\section*{Code and data availability}

Scripts for VASP MD simulations and all analysis discussed in this publication are available on Github \cite{strachangroup-dopedllzo}.A reduced example for a system is also provided there.

\section*{Acknowledgements}

We acknowledge computational resources from nanoHUB and Purdue University through the Network for Computational Nanotechnology. J. C. V. thanks the Science and Technology Council of Mexico (Consejo Nacional de Ciencia y Tecnología, CONACYT) for its financial support of this research. The authors gratefully acknowledge valuable comments from Dr. Omar Israel Gonzalez-Pena from the Monterrey Institute of Technology and Higher Education.

\section*{Ethics declarations}

\subsection*{Conflict of interest}

On behalf of all authors, the corresponding author states that there is no conflict of interest.

\bibliographystyle{unsrt}
\bibliography{references.bib}

\section*{Supporting Information}

Supplementary figures and tables related to the analysis 

\pagebreak



\renewcommand{\thefigure}{S\arabic{figure}}
\renewcommand{\thetable}{S\arabic{table}}
\renewcommand{\thepage}{S\arabic{page}}
\setcounter{figure}{0}
\setcounter{table}{0}
\setcounter{page}{1}

\title{Supporting Information: Atomistic mechanisms underlying the maximum in diffusivity in doped Li$_7$La$_3$Zr$_2$O$_{12}$}

\maketitle

\begin{figure*}[ht]
\centering
\begin{subfigure}{.33\textwidth}
\centering
\subcaptionOverlay{\includegraphics[width=\textwidth]{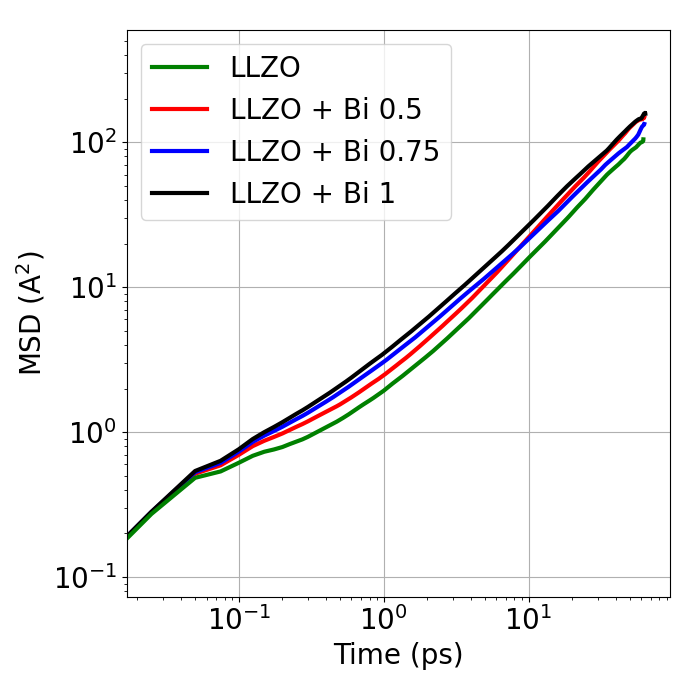}}
\end{subfigure}%
\begin{subfigure}{.33\textwidth}
  \centering
\subcaptionOverlay{\includegraphics[width=\textwidth]{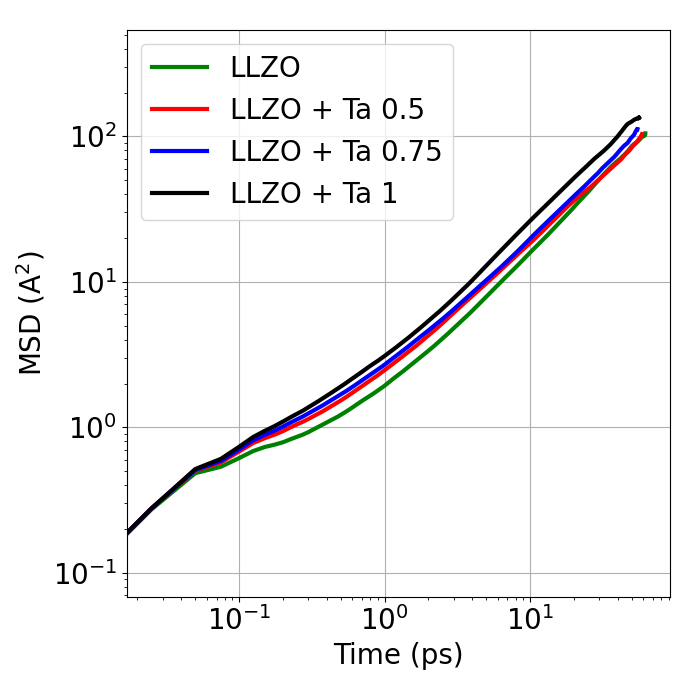}}
\end{subfigure}
\begin{subfigure}{.33\textwidth}
  \centering
  \subcaptionOverlay{\includegraphics[width=\textwidth]{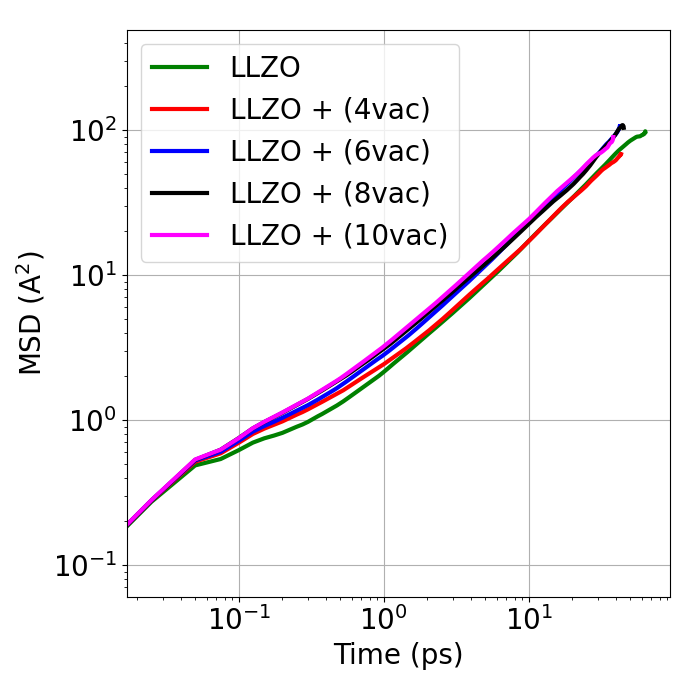}}
\end{subfigure}
\caption{Time average mean square displacement (MSD) for LLZO structures at 1773K, (a) Bi-doped structures, (b) Ta-doped structures, and (c) structures with artificial vacancies.}
\label{rawMSD}
\end{figure*}


\begin{table}[ht]
\caption{Transitions between tetrahedral and octahedral sites for each system.}
\label{transitiontable}
\vskip 0.15 in
\begin{center}
\begin{small}
\begin{sc}
\begin{tabular}{lccccr}
\toprule
Structure	&	\makecell{T $\rightarrow$ O \\ Jumps \\ (1773K)}	&	\makecell{O $\rightarrow$ T \\Jumps\\ (1773K)}	&	\makecell{T $\rightarrow$ O \\ Jumps \\ (1273K)}	&	\makecell{O $\rightarrow$ T \\ Jumps \\(1273K)}	\\
\midrule
LLZO	&	731	&	733	&	475	&	474	\\
LLZO (Bi = 0.5)	&	896	&	900	&	787	&	785	\\
LLZO (Bi = 0.75)	&	1074	&	1079	&	792	&	793	\\
LLZO (Bi = 1)	&	1179	&	1187	&	918	&	918	\\
LLZO (Ta = 0.5)	&	850	&	849	&	756	&	753	\\
LLZO (Ta = 0.75)	&	837	&	835	&	803	&	801	\\
LLZO (Ta = 1)	&	946	&	948	&	790	&	786	\\
LLZO (LiV = 0.5)	&	645	&	639	&	521	&	521	\\
LLZO (LiV = 0.75)	&	669	&	673	&	341	&	337	\\
LLZO (LiV = 1)	&	768	&	773	&	574	&	575	\\
LLZO (LiV = 1.25)	&	657	&	658	&	504	&	512	\\
\bottomrule
\end{tabular}
\end{sc}
\end{small}
\end{center}
\vskip -0.1in
\end{table}

\begin{figure}[h]
    \centering
    \includegraphics[width=\textwidth]{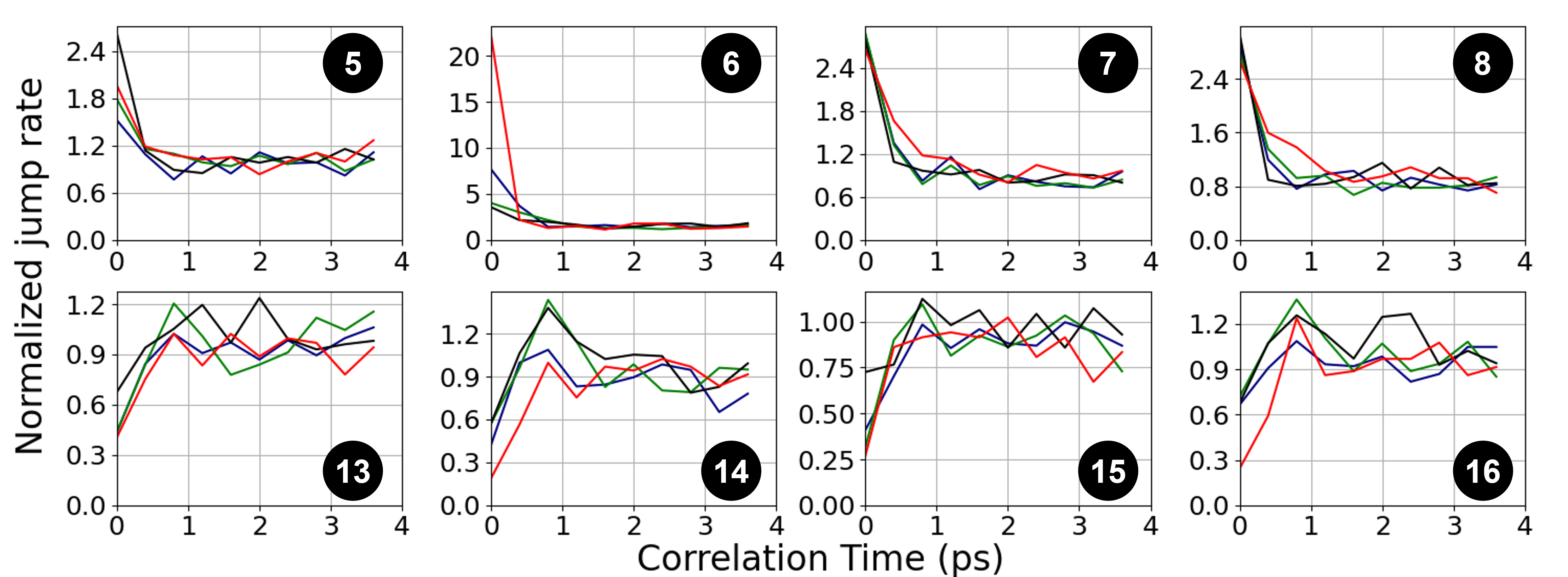}
    \caption{Normalized jumping rate against correlation time for selected mechanisms for the family of Bi-doped LLZO structures at 1273K. Colors represent different dopant contents. [Red = 0, Blue = 0.5, Green = 0.75, Black = 1]. Table \ref{pairwisetable} describes of the pair correlation cases.}
    \label{supp:lowtemp_norm_corr_plots}
\end{figure}
\pagebreak

\begin{figure*}[ht]
\centering
\begin{subfigure}{.4\textwidth}
\centering
\subcaptionOverlay{\includegraphics[width=\textwidth]{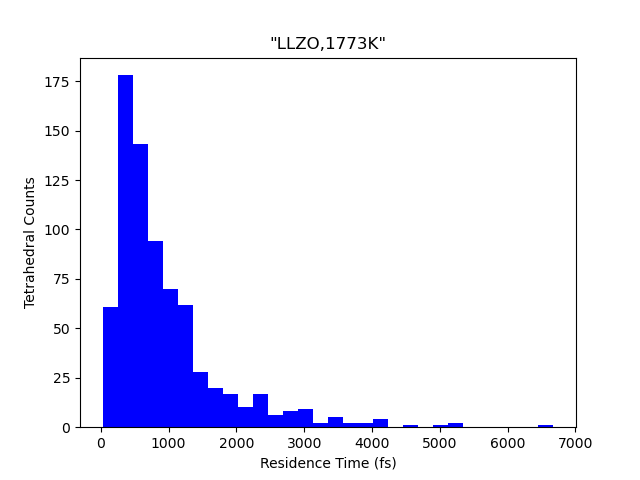}}
\end{subfigure}
\begin{subfigure}{.4\textwidth}
\centering
\subcaptionOverlay{\includegraphics[width=\textwidth]{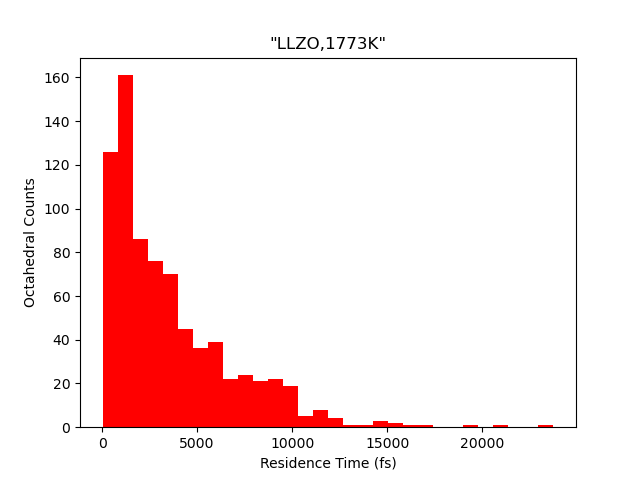}}
\end{subfigure}
\begin{subfigure}{.4\textwidth}
\centering
\subcaptionOverlay{\includegraphics[width=\textwidth]{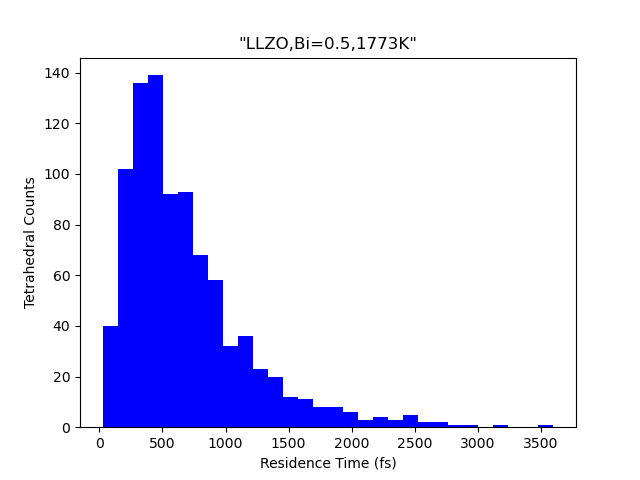}}
\end{subfigure}
\begin{subfigure}{.4\textwidth}
\centering
\subcaptionOverlay{\includegraphics[width=\textwidth]{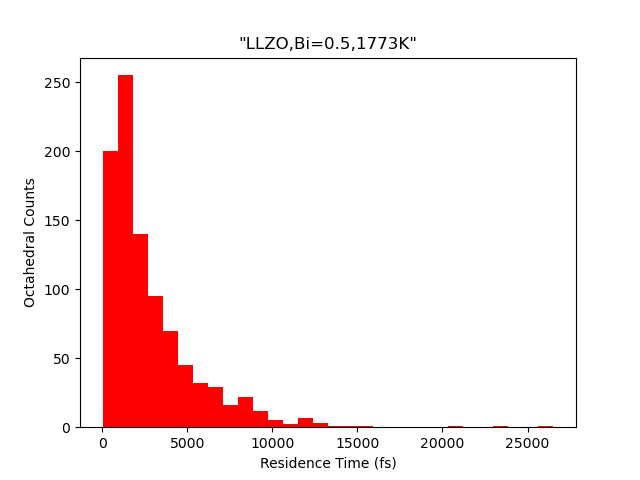}}
\end{subfigure}
\begin{subfigure}{.4\textwidth}
\centering
\subcaptionOverlay{\includegraphics[width=\textwidth]{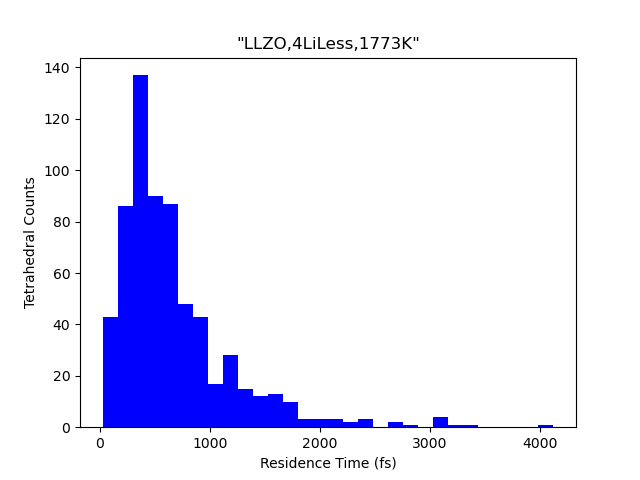}}
\end{subfigure}
\begin{subfigure}{.4\textwidth}
\centering
\subcaptionOverlay{\includegraphics[width=\textwidth]{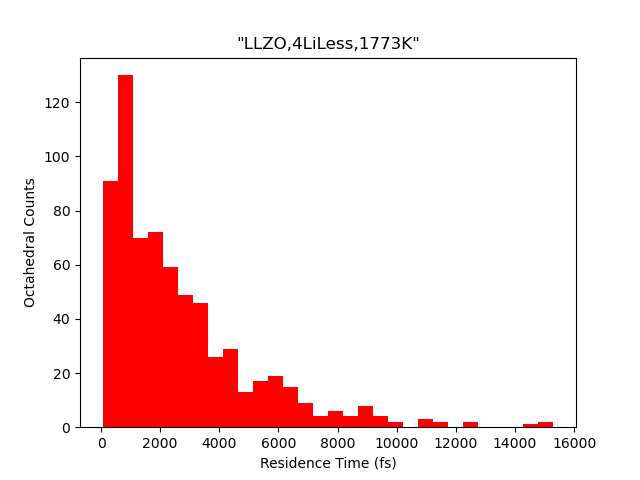}}
\end{subfigure}
\begin{subfigure}{.4\textwidth}
\centering
\subcaptionOverlay{\includegraphics[width=\textwidth]{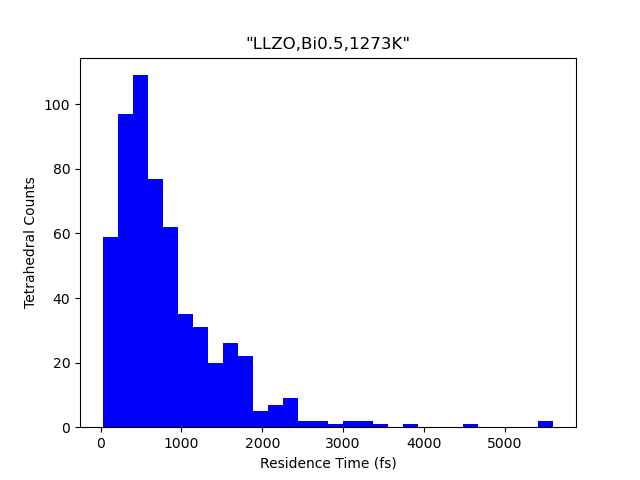}}
\end{subfigure}
\begin{subfigure}{.4\textwidth}
\centering
\subcaptionOverlay{\includegraphics[width=\textwidth]{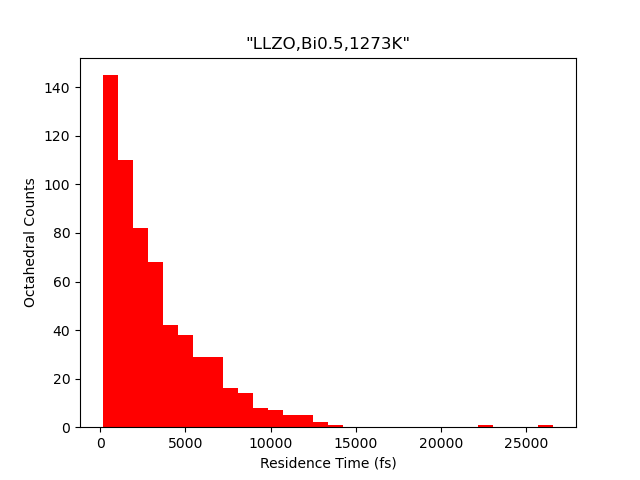}}
\end{subfigure}
\caption{Residence time distributions for (a,b) LLZO (1773K), (c-d) LLZO (Bi = 0.5) (1773K), (e-f) LLZO (LiV = 0.5)(1773K), (g-h) LLZO (Bi = 0.5) (1273K).}
\label{jumpstats_figs3}
\end{figure*}


\begin{table}[h]
\caption{Jumping statistics for residence times of doped LLZO garnets.}
\label{rtime_statisticstable}
\vskip 0.15 in
\begin{center}
\begin{small}
\begin{sc}
\begin{tabular}{lccccr}
\toprule
Structure	&	\makecell{Tetrahedral \\ Res Time \\ (Mean) (ps) }	& \makecell{Tetrahedral \\ Res Time \\ (95\% CI) (ps)}	&\makecell{Octahedral \\ Res Time \\ (Mean) (ps)}	&	\makecell{Octahedral \\ Res Time \\ (95\% CI) (ps)} \\ 
\midrule

LLZO (1773K)	&	0.930	&	0.059	&	3.586	&	0.231	\\
LLZO (Bi = 0.5) (1773K)	&	0.683	&	0.033	&	2.817	&	0.176	\\
LLZO (LiV = 0.5)(1773K) &	0.669	&	0.041	&	2.656	&	0.182	\\
LLZO (Bi = 0.5) (1273K)	&	0.828	&	0.056	&	3.266	&	0.238	\\

\bottomrule
\end{tabular}
\end{sc}
\end{small}
\end{center}
\vskip -0.1in
\end{table}

\end{document}